\documentclass{PoS}
\usepackage{float}
\usepackage[caption = false]{subfig}
\usepackage{sidecap}
\usepackage{lineno}

\title{Probing Nuclear PDF and Gluon Saturation At The LHC with Forward Direct Photons}

\ShortTitle{Probing nPDF and Gluon Saturation with Forward Direct Photons}

\author{\speaker{Mauro R. Cosentino}, for the ALICE FoCal Collaboration\\
        Centro de Ciencias Naturais e Humanas, Universidade Federal do ABC\\
        CEP 09210-580, Santo Andre, SP, Brazil\\
        E-mail: \email{mcosenti@cern.ch}}

\abstract{In relativistic nuclear collisions some of the important aspects to be addressed are the effects of the nuclear PDF and the gluon saturation. In the LHC the best way to address these questions is by means of pA collisions and in particular through the measurement of direct photon production in the forward direction. In order to achieve this measurement a new forward calorimeter (FoCal) is proposed as an upgrade to the ALICE experiment. The proposed detector will cover the range 3.5 < $\eta$ < 5, probing the gluon distributions at x$\sim$10$^{-5}$ and Q$\sim p_{\rm T}$ > 4 GeV.  We will discuss performance studies and demonstrate that extremely high-granularity calorimetry is required for such measurement. We will also present a few results from R\&D for this project.}

\FullConference{The 26th International Nuclear Physics Conference\\
		11-16 September, 2016\\
		Adelaide, Australia}

\begin{document}

\section{Motivation}
\label{sec:motiv}

The inner structure of hadrons, known as Parton Distribution Functions (PDFs) for protons and neutrons and as nPDFs for nuclei, is of fundamental importance in the understanding of the physics results coming from high energy hadron colliders, such as the LHC. Since the PDFs are  initial state configurations, they can impact many of final observables like cross sections \cite{Martin} or azimuthal angular correlations \cite{DrescherNara}. 

\begin{figure}
	\centering
	\subfloat{\includegraphics[width=.32\textwidth]{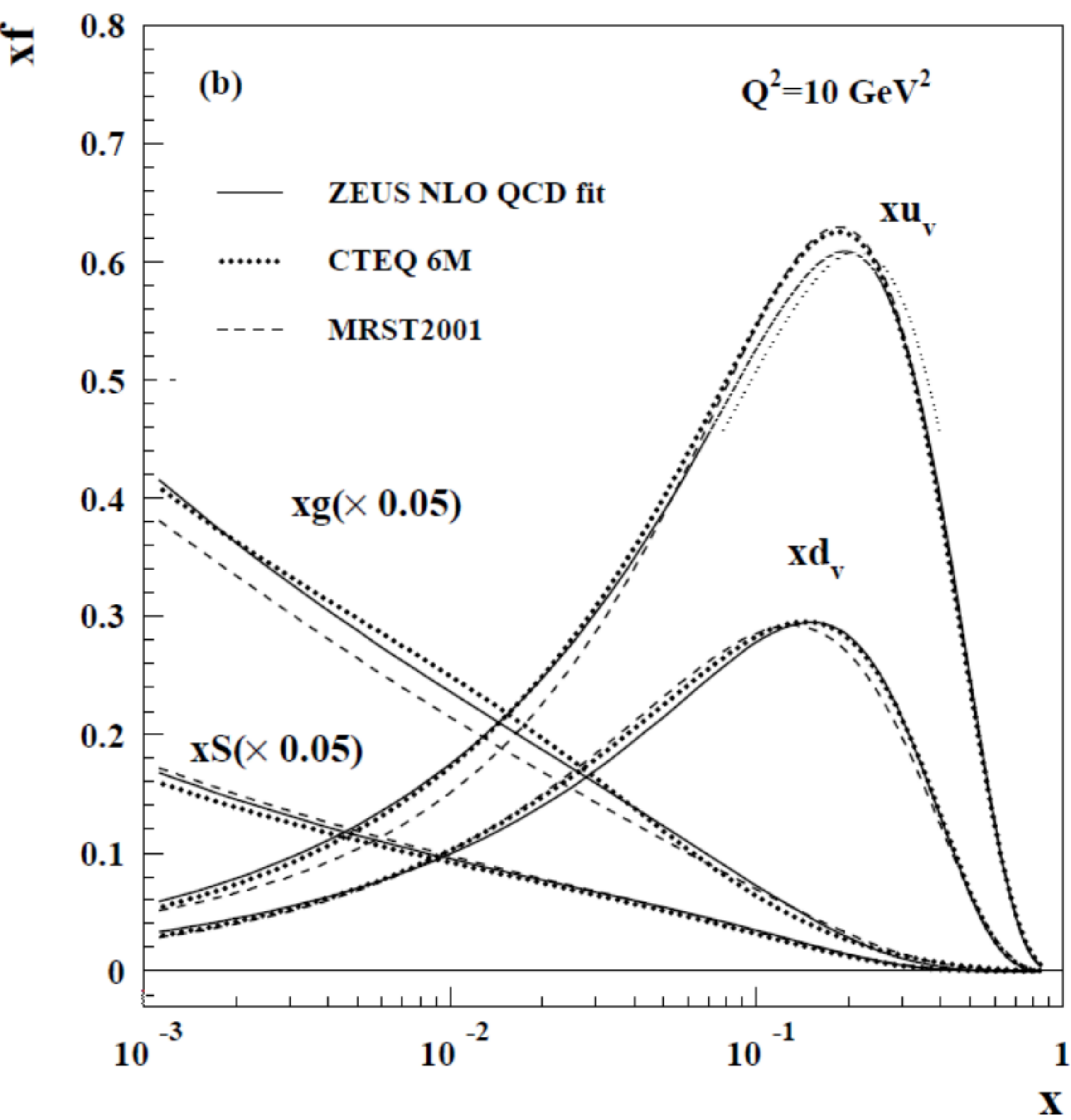}}
	\hspace{0.10\textwidth}
	\subfloat{\includegraphics[width=.4\textwidth]{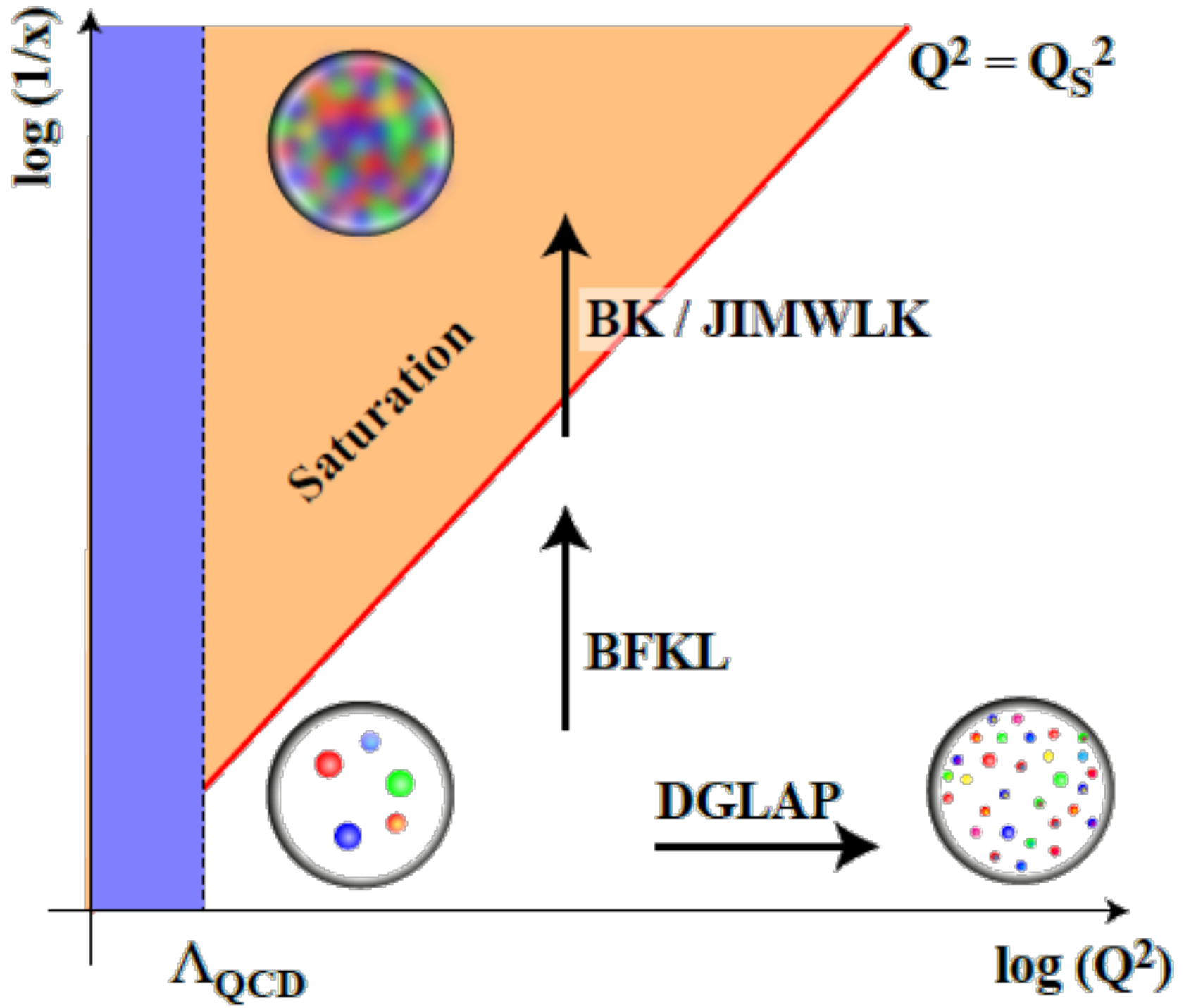}}
	\caption{Evolution of PDFs. Left: proton PDF for various fits. (Gluon and sea-quark distributions - xg and xS - are scaled down by a factor of 20\cite{ZeusPDF}). Right: the gluon saturation diagram according to evolution equations.}
\label{fig:fig1}
\end{figure}

In addition to the above, nPDFs are of great interest themselves, given the fact that the DGLAP equations, that describe their evolution, predict a divergence of the gluon densities for decreasing values of longitudinal momentum fraction $x$, violating, for example, unitarity. Fig. \ref{fig:fig1} (left), presents the PDFs for the proton as obtained from Deep Inelastic Scattering (DIS) from ZEUS data \cite{ZeusPDF}, showing the gluon density tendency to diverge. The gluon density - schematically presented in the right panel of Fig. \ref{fig:fig1} - increases for decreasing $x$ and increasing momentum transfer, $Q^2$, up to the point where non-linearity effects are expected to become important and a saturation of gluons is expected to limit the gluon distribution. Gluon saturation is described by models such as the {\it Color Glass Condensate} (CGC) \cite{CGC}, which describe quantitatively the initial states of highly energetic protons and nuclei.

A saturation scale for the parton distribution of gluons can be estimated by
\begin{equation}\label{eq:q2sat}
Q_s^2(x)\approx\frac{\alpha_S}{\pi R^2}xG(x,Q^2)\propto \left(\frac{A}{x}\right)^{1/3}
\end{equation}
showing that for a given value of $x$, the heavier the nucleus\footnote{The $A$ dependence comes from $G\propto A$ and $R\propto A^{1/3}$}, the larger is $Q_s^2$, making this a more accessible measurement. In addition to this, the minimum accessible value of $x$ can be estimated as a function of the rapidity $y$ and the collision energy by\footnote{From an approximation of a LO calculation of a 2$\rightarrow$2 process.}

\begin{equation}\label{eq:xmin}
x_{min}\approx\frac{2p_{\rm T}}{\sqrt{s_{NN}}}e^{-y}
\end{equation}

Hence, equations \ref{eq:q2sat} and \ref{eq:xmin} show that saturation can be more easily probed in collisions of protons on heavy nuclei targets with measurements taken in the forward region. The higher collision energy of the LHC provides access to x as low as $x\sim10^{-5}$ at $\langle y\rangle\approx$4. The use of protons as projectiles has the purpose of avoiding the final state effects present in $A-A$ collisions.

\subsection{Saturation I: Indications from RHIC}
\label{subsec:rhic-results}

\begin{figure}
	\subfloat{\includegraphics[width=.3\textwidth]{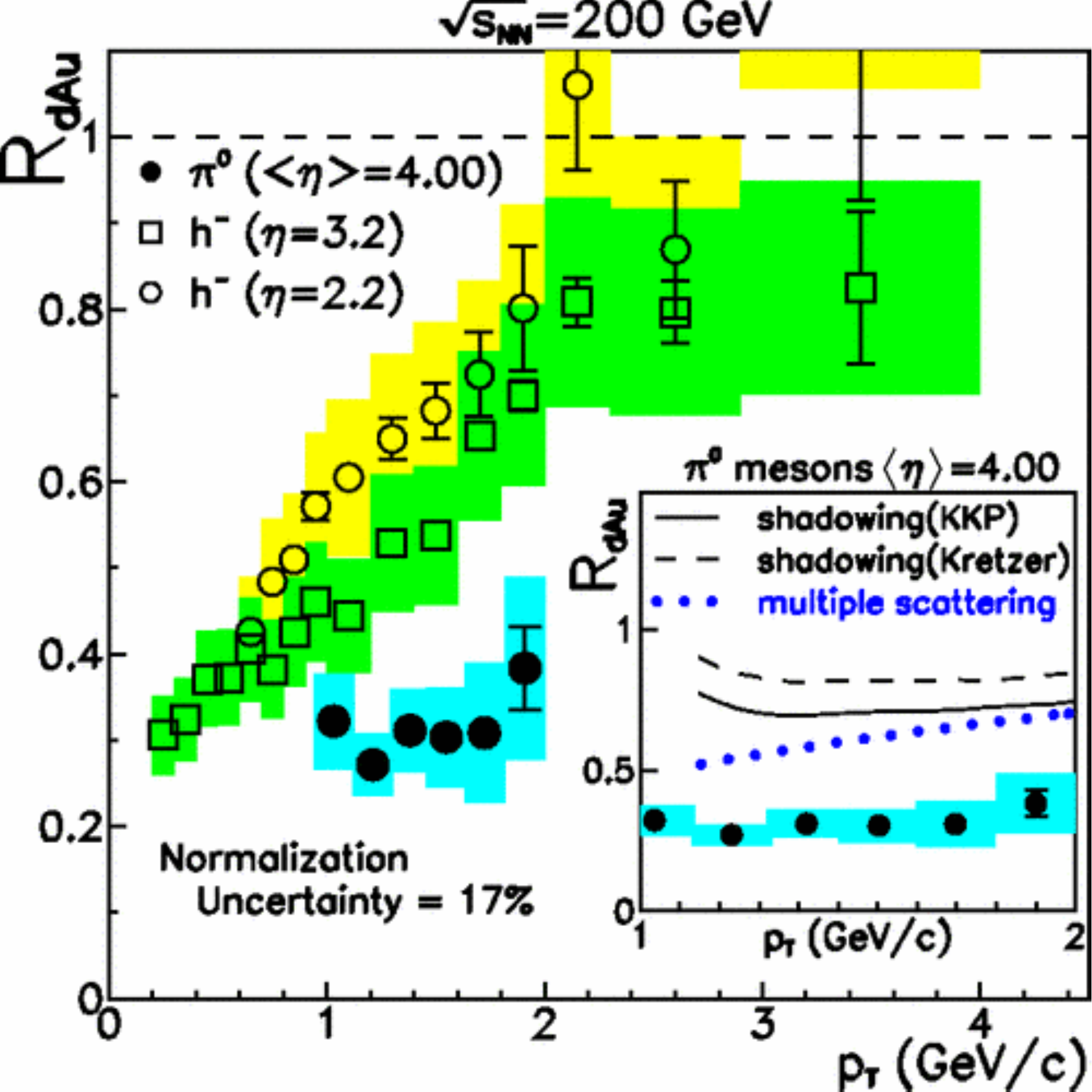}}
	\hspace{0.05\textwidth}
	\subfloat{\includegraphics[width=.6\textwidth]{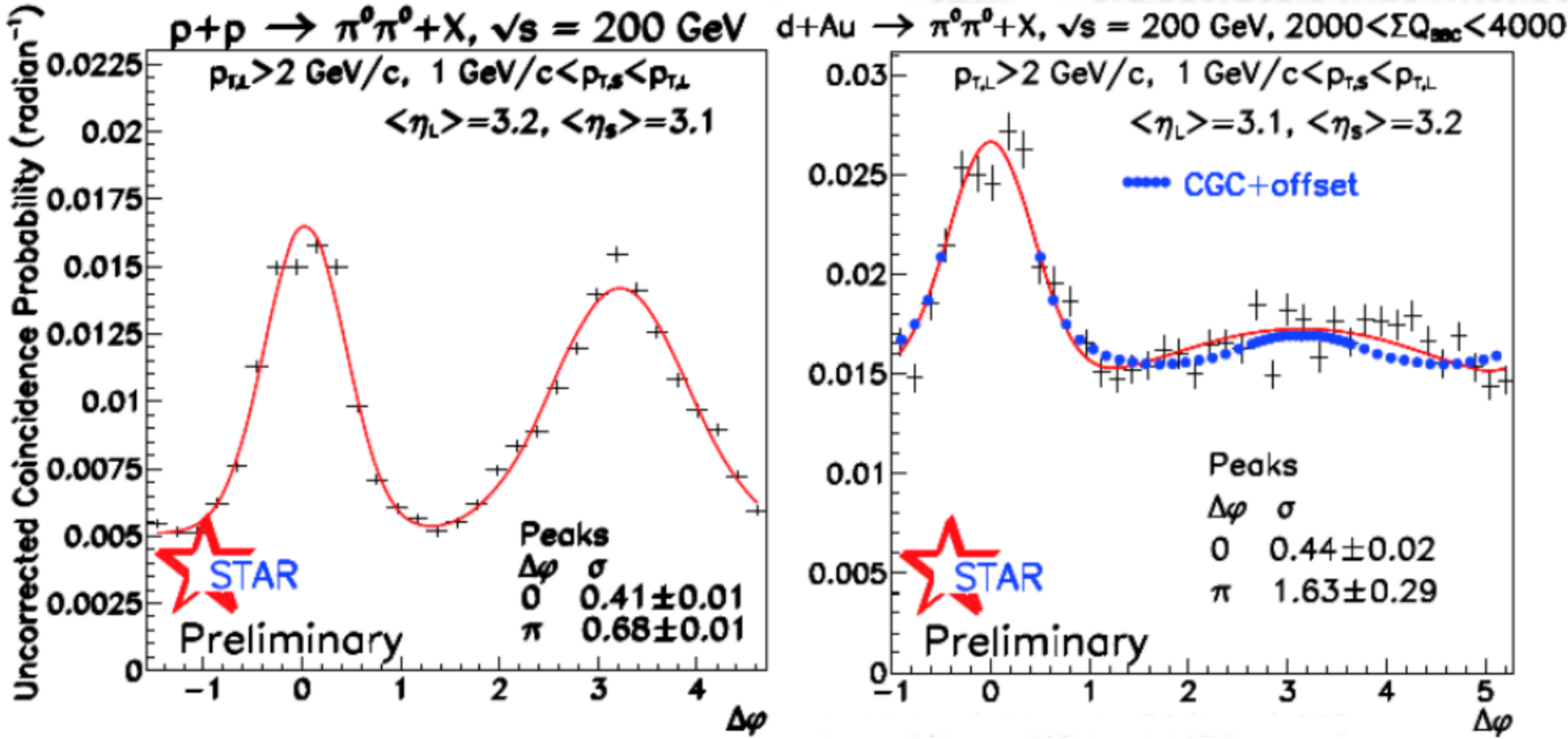}}
	\caption{RHIC d+Au results. Left: $R_{dAu}$ $\eta$ dependency, showing an increasing suppression pattern. Right: Comparison of two hadron azimuthal correlations in p+p and d+Au, where a significant suppression of the away side is observed.\cite{RHIC}}
\label{fig:rhic}
\end{figure}

At the Relativistic Heavy Ion Collider (RHIC), measurements aimed to probe gluon saturation are performed in d-Au collisions. At RHIC, two of the most striking results that support the CGC model are the ones presented in Fig. \ref{fig:rhic}. On the left the inclusive hadron $p_{\rm T}$ spectra scaled by the equivalent number of proton-proton collisions given by the variable 
\begin{equation}\label{eq:rdau}
R_{dAu} = \frac{d^2N/(d\eta dp_{\rm T})\mid_{dAu}}{\langle N_{coll}(dAu)\rangle d^2N/(d\eta dp_{\rm T})\mid_{pp}}
\end{equation}
shows a much stronger suppression in the forward region ($\langle\eta\rangle$=4) than in more central rapidity values, with the overall picture showing a pattern of increasing suppression with rapidity.

The other important result is the one on the right of Fig. \ref{fig:rhic}, where the azimuthal angular distribution, with respect to a trigger particle, of neutral pions measured in the forward region is compared between the two colliding systems - $pp$ and d-Au. There is a strong suppression of the away side ($\Delta\phi\approx\pi$) in the d-Au collisions with small impact parameter, when the deuteron hits the $Au$ nucleus where the nucleon - and hence parton - density is higher.
Both results are consistent with CGC predictions - that predicts that the recoil parton is in a coherent state with many others - but other hadronic final state interactions, such as hadronic absorption, inital state energy loss and/or multiple scatterings effect cannot be ruled out. 

\subsection{Saturation II: Indications from LHC}
\label{subsec:lhc-results}

\begin{figure}
	\centering
	\subfloat{\includegraphics[width=.355\textwidth]{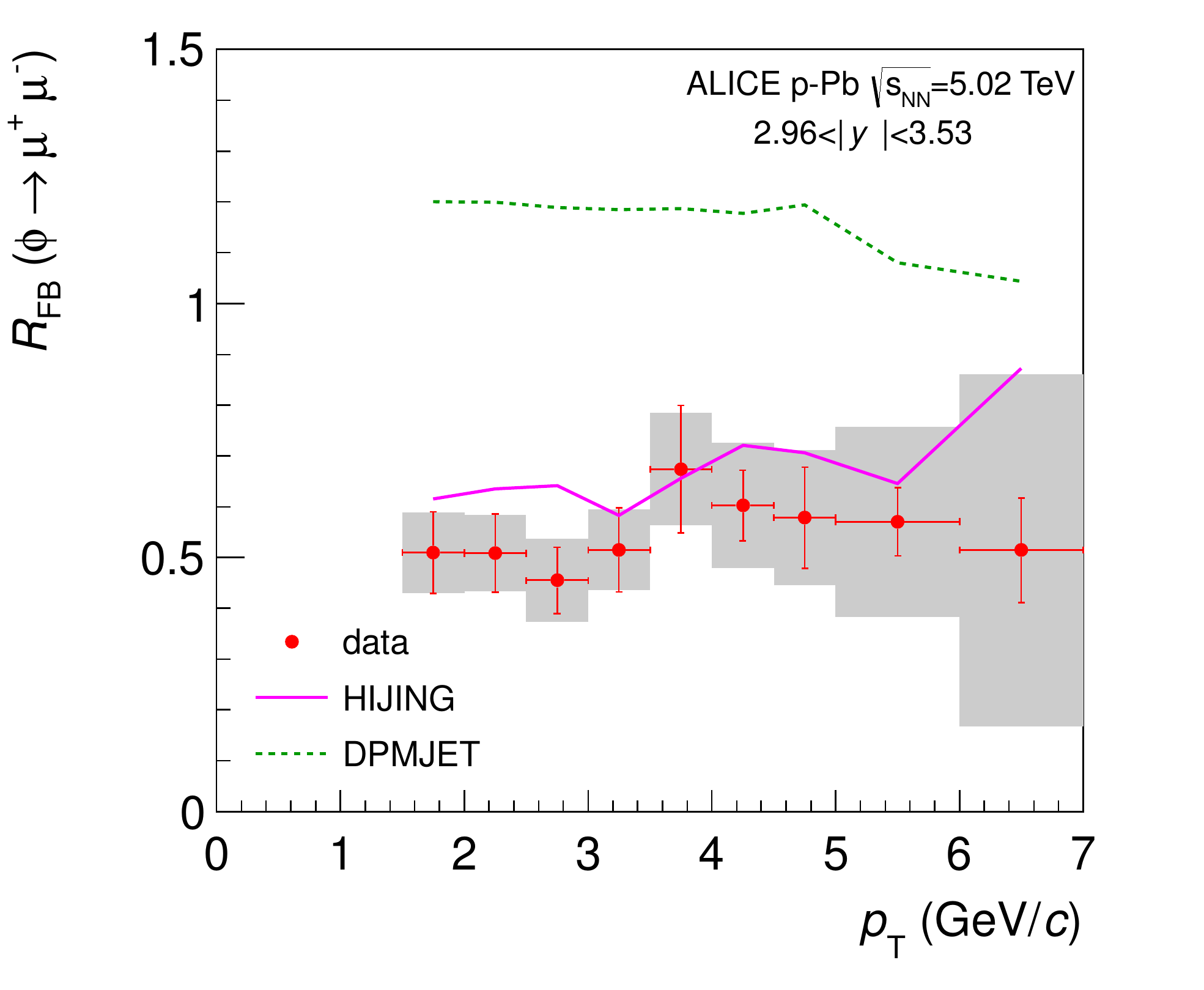}}
	\hspace{0.03\textwidth}
	\subfloat{\includegraphics[width=.43\textwidth]{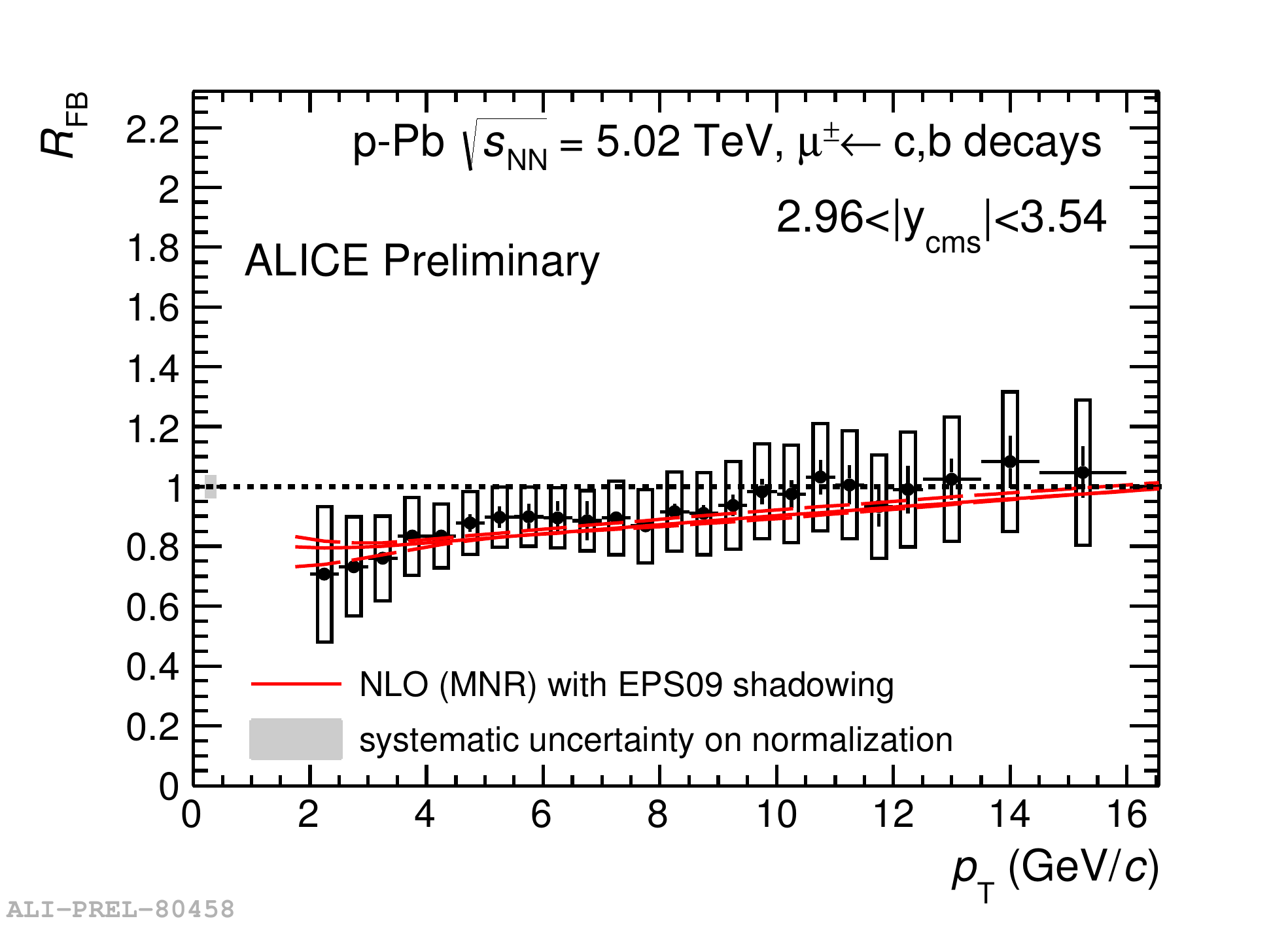}}
	\caption{ALICE $R_{FB}$ results in p--Pb in the LHC. Left: $R_{FB}$ for $\phi$ mesons showing a strong suppression. Right: $R_{FB}$ for $\mu\leftarrow$c,b, presenting a slight suppression at low $p_{\rm T}$.\cite{LHC}}
\label{fig:lhc}
\end{figure}

At the LHC the scenario is similar to the one at RHIC in the sense that there are several results presenting modification in p--Pb collisions with respect to the same observables in $pp$ that may be the manifestation of initial state effects, but that have many other possible hadronic effects contributing to them. Fig. \ref{fig:lhc} presents some of the ALICE experiment p--Pb results where the comparison to $pp$ is avoided by the creation of an observable based on the {\it Forward-to-Backward Ratio} or $R_{FB}$, which is simply the yield measured in forward (proton going) divided by the yield in the backward (lead going) direction, within the common rapidity coverage of both.

On the left side of Fig. \ref{fig:lhc}, the $R_{FB}$ of $\phi$ mesons shows a strong suppression for which the physical mechanism is not yet clear. On the right side of the figure, the $R_{FB}$ of muon from charm and bottom decays is consistent with shadowing. However the initial state kinematics ($x$, $Q^2$) of heavy flavour  are not well defined, making these results no yet very conclusive.

In summary we may say that in $p(d)-A$ collisions, there are striking results at RHIC, and several results at the LHC showing suppression of hadron production in processes that are sensitive to low-$x$ gluon density, but all of them are based in hadronic probes that are subject to many other sources of influence other than the initial parton distribution. The natural alternative to this is to make use of electromagnetic probes, that can be sensitive to the gluon parton distribution and be free of undesired final state dynamics.

\subsection{Probing Saturation with Direct Photons}
\label{subsec:dirpho}

\begin{figure}
	\hspace{0.10\textwidth}
	\subfloat{\includegraphics[width=.40\textwidth]{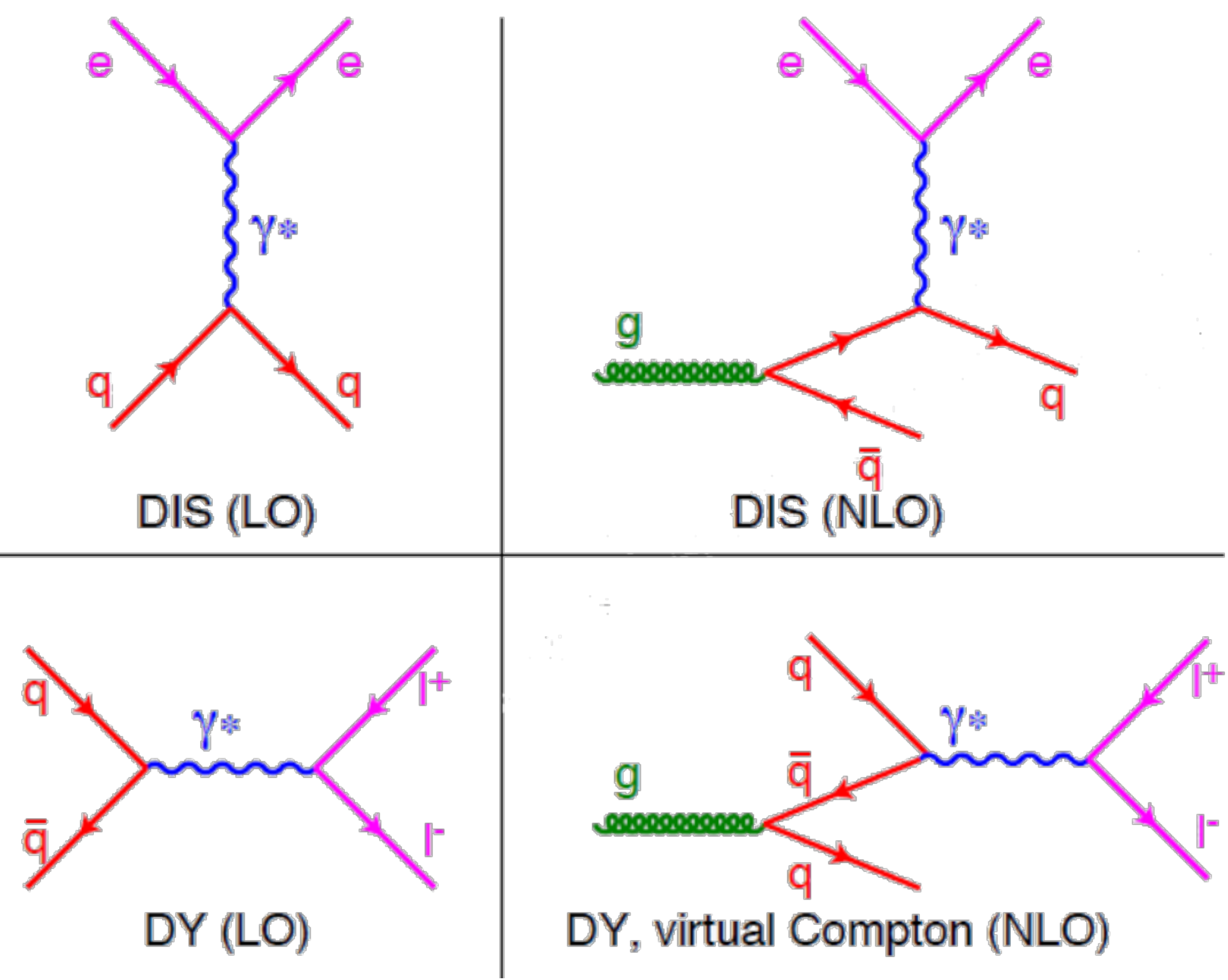}}
	\hspace{0.15\textwidth}
	\subfloat{\includegraphics[width=.30\textwidth]{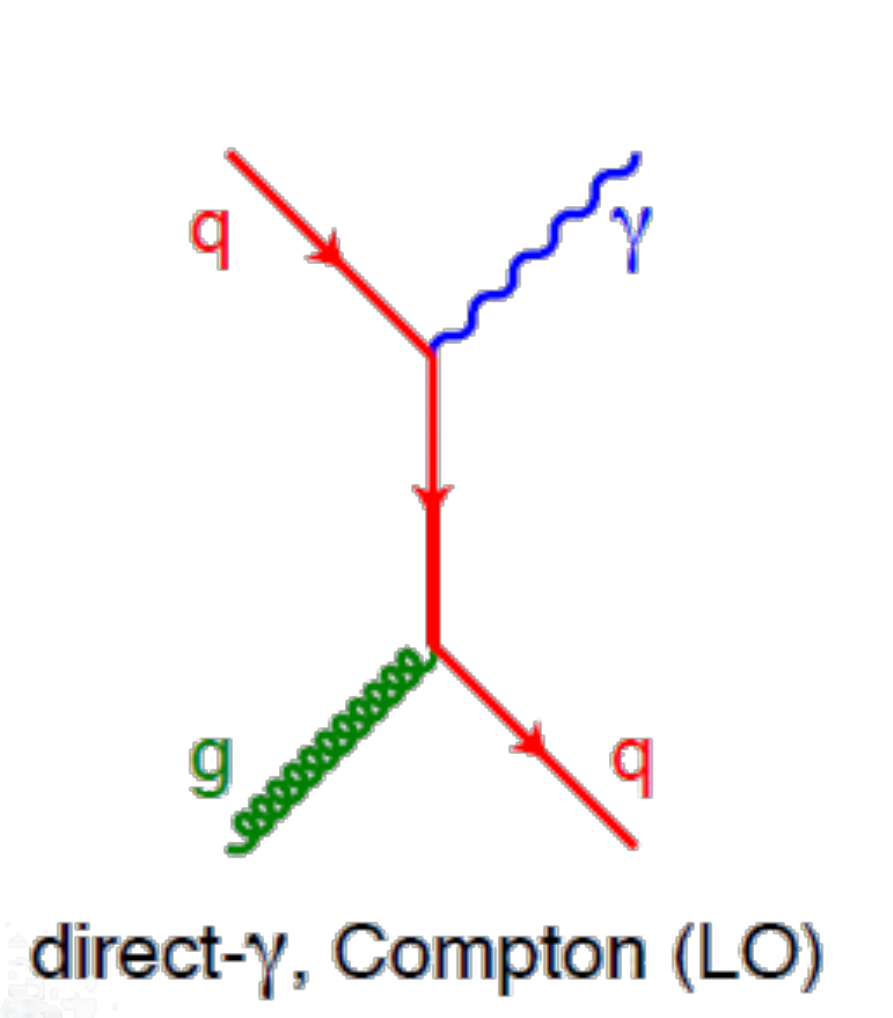}}
	\caption{Feynman diagrams for the production of different electromagnetic probes. Left: Drell-Yan processes and its correspondence to DIS, showing their gluon sensitivity only at NLO. Right: Direct photon production in a "Compton like" diagram which is sensitive to gluons already at LO.}
\label{fig:diagrams}
\end{figure}

Dileptons from Drell-Yan (DY) processes can be used to probe the nPDFs in the very same way as in DIS, since they are equivalent processes, as shown by the Feynman diagrams on the left of Fig. \ref{fig:diagrams}, where the scattering electron in DIS corresponds to the outgoing (anti)lepton pair. From those one can see that, in fact, DY is sensitive to the gluon distributions only at NLO, which decreases significantly the cross section of the desired processes in one hand and on the other the LO process guarantees that most of the dileptons will be background.

The case for direct photons is more promising. The right side of Fig. \ref{fig:diagrams} shows the "Compton-like" diagram which is a LO process sensitive to the gluon parton distribution function. The expected luminosity of the future LHC p--Pb runs may not allow a DY measurement, while the direct photon, with its higher cross section, can be a promising probe.

\section{The ALICE experiment and the FoCal as an Upgrade}
\label{subsec:alice}

The measurement of direct photons at forward rapidity in pp and p--Pb collisions at LHC is the main physics motivation for the proposed FoCal upgrade of the ALICE experiment. The proposal is presently under discussion within the Collaboration and would be targeted for installation during the third long shutdown of the LHC in 2024.

\begin{figure}
	\subfloat{\includegraphics[width=.5\textwidth]{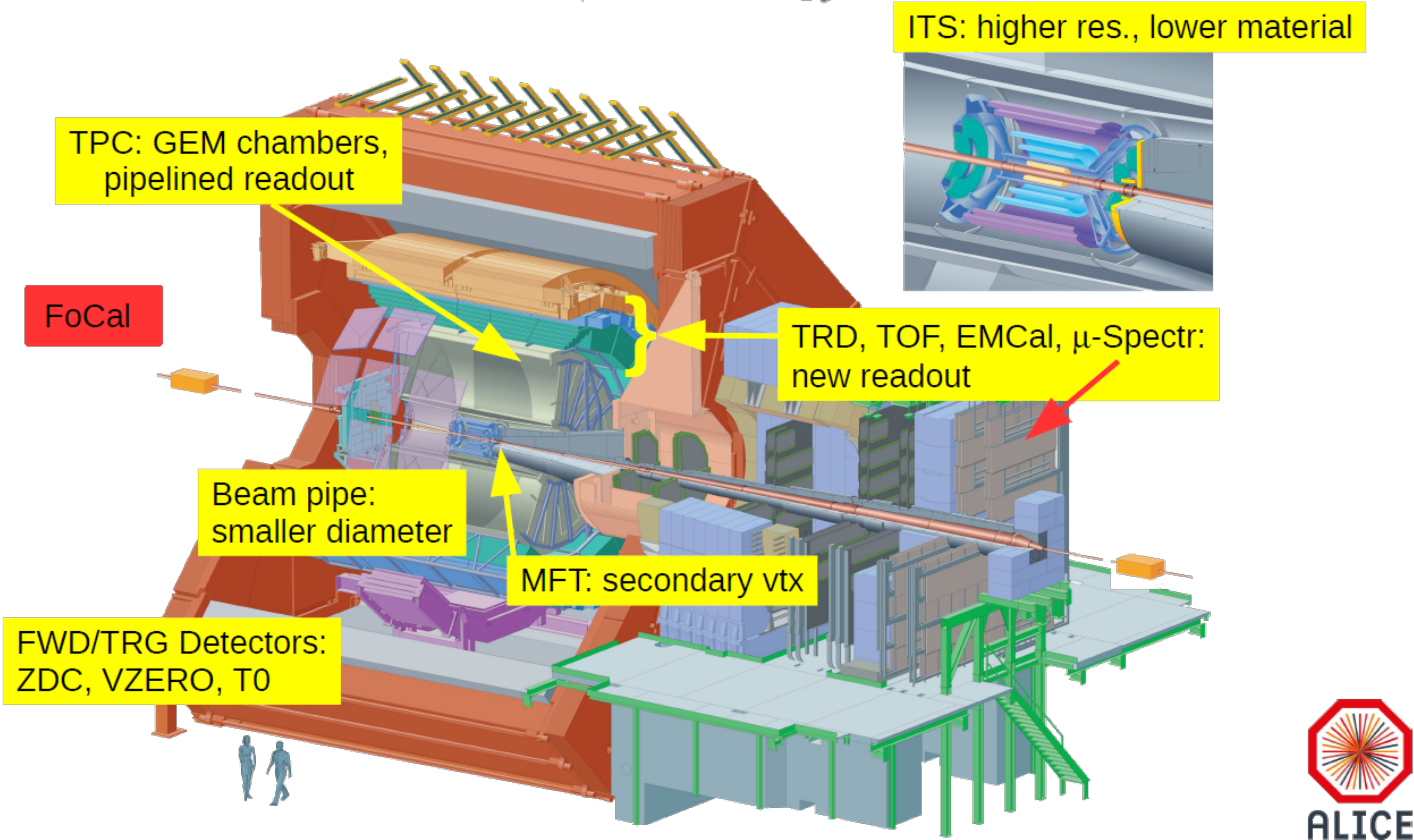}}
	\hspace{0.05\textwidth}
	\subfloat{\includegraphics[width=.5\textwidth]{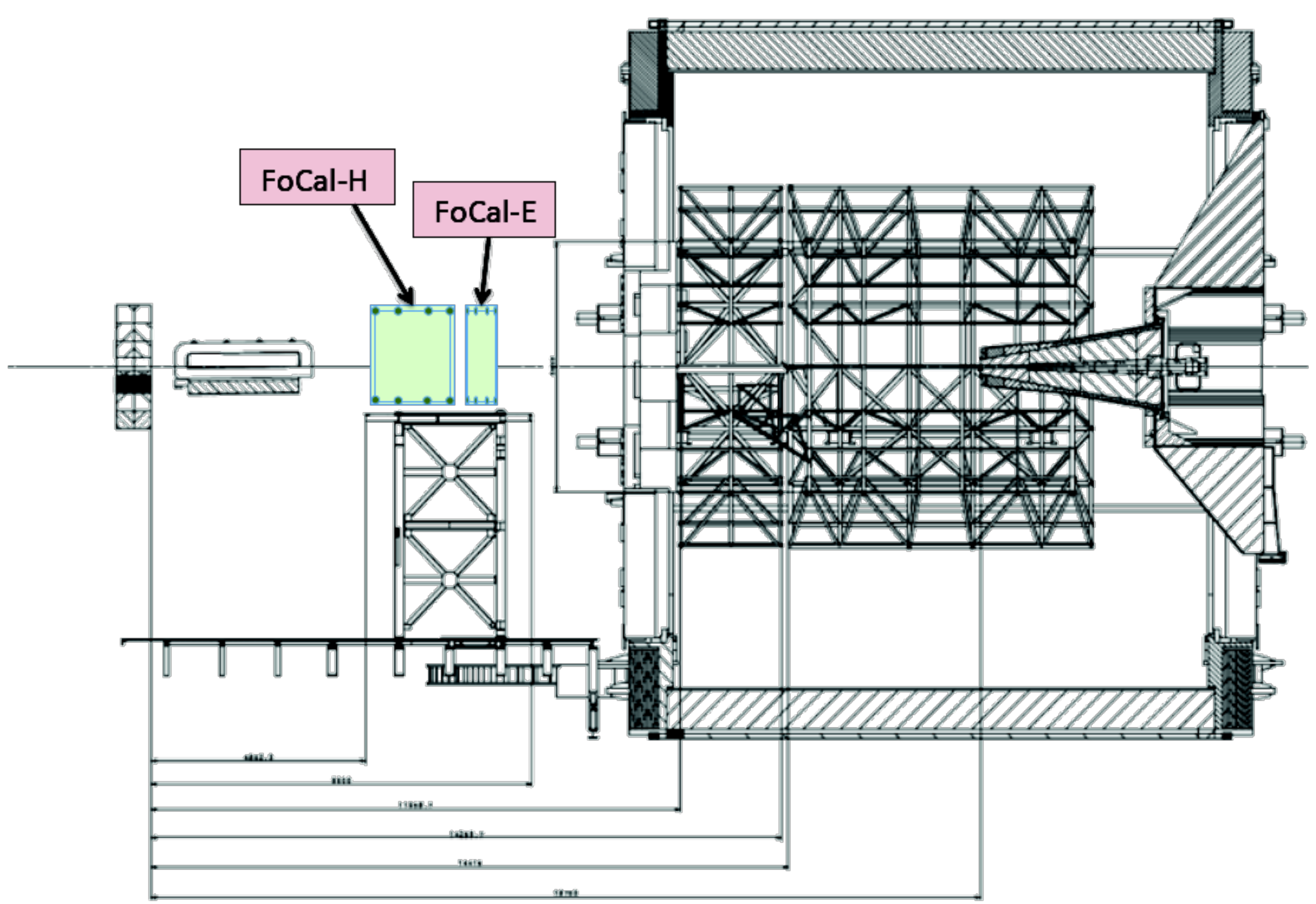}}
	\caption{Schematic view of the ALICE experiment. Left: current ALICE configuration with the indication of the upgrades. Right: longitudinal view of the ALICE experiment and the placement for the FoCal.}
\label{fig:setup}
\end{figure}

The current setup of ALICE \cite{aliperf} is shown schematically in Fig. \ref{fig:setup}. On the left is presented the overview of the experiment with labels showing the places of the approved upgrades\cite{aliupgrade}, and the proposed position for the FoCal (in red). The right side of Fig.\ref{fig:setup} presents the side-view of the experiment with the possible location for the two parts of the FoCal: the electromagnetic and hadronic calorimeters.

The FoCal project aims for a calorimeter for $\gamma$ and $\pi^0$ measurements with enough resolution to reconstruct $\pi^0$s either by the invariant mass of the two $\gamma$ showers reconstructed separately or through the shape of the reconstruct overlapped showers. To cover the largest rapidity possible, the FoCal is placed at $\sim$8 m from the interaction point, covering 3.3<$\eta$<5.3, where $\pi^0$s energy are high and their decays opening angle is small. Hence, the Moli\`ere radius ($R_M$) should be as small as possible, to enable very high spatial resolution. For that it also needs a very high granularity read-out.

\begin{figure}
	\subfloat{\includegraphics[width=.5\textwidth]{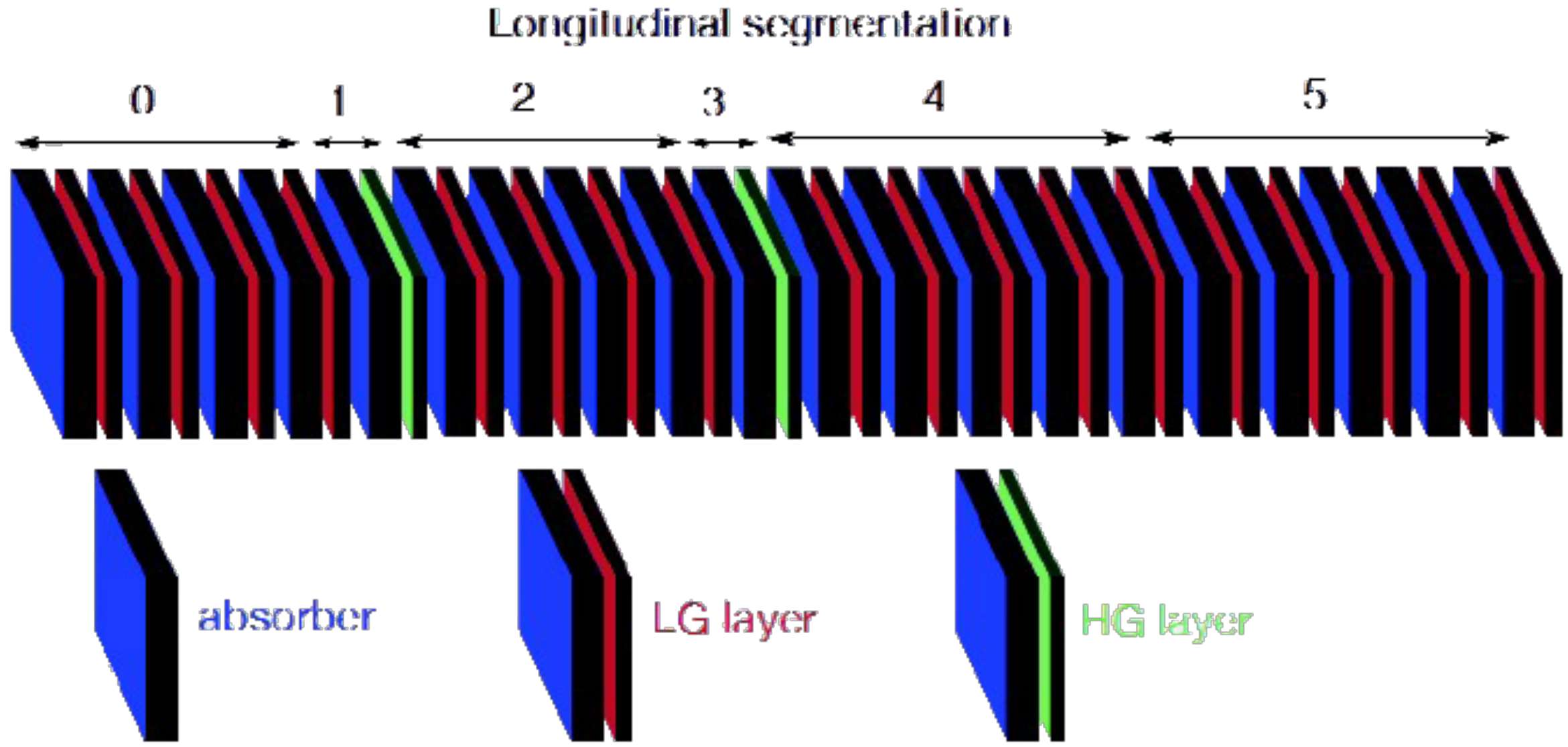}}
	\hspace{0.05\textwidth}
	\subfloat{\includegraphics[width=.5\textwidth]{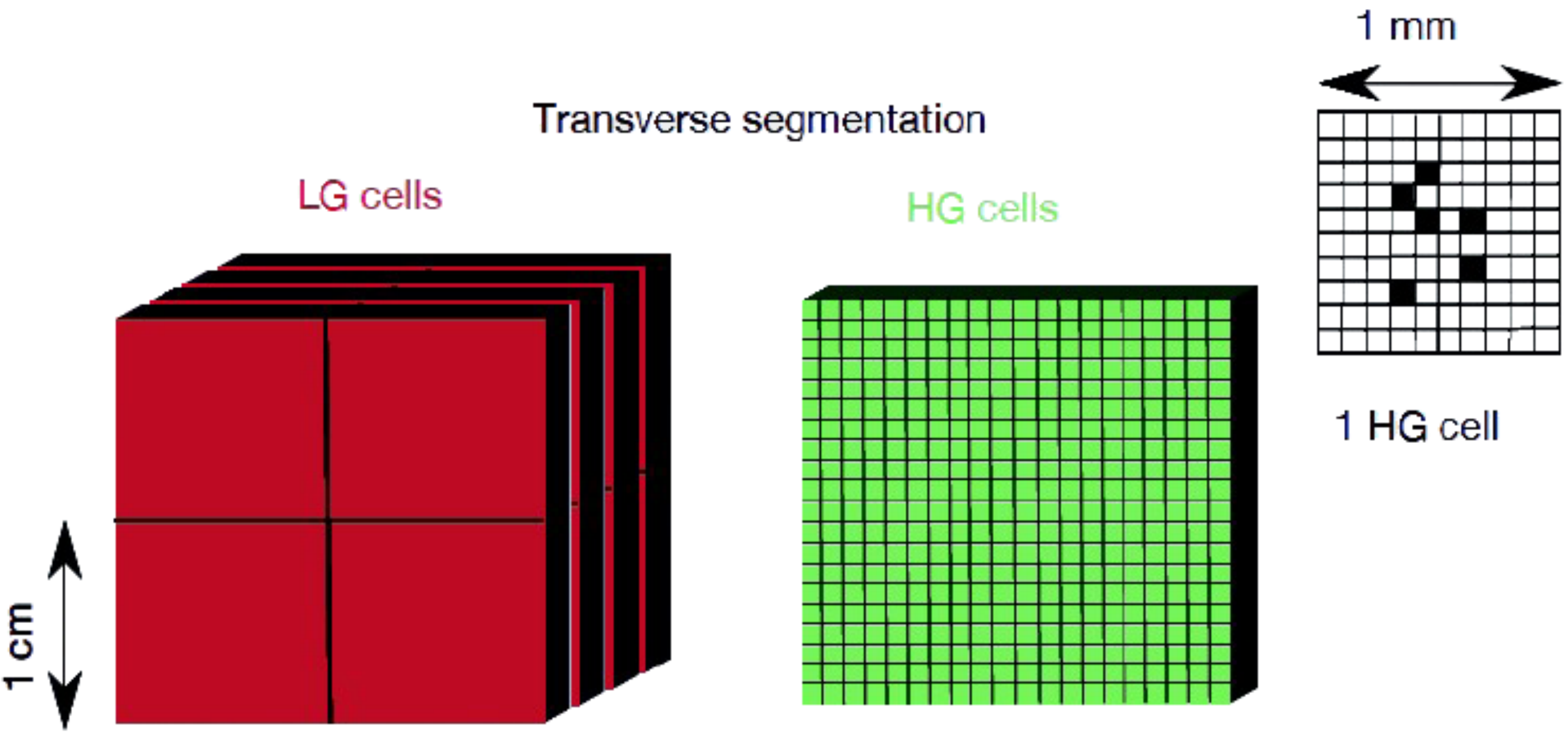}}
	\caption{The FoCal strawman design. Left: the longitudinal segmentation, divided in layers of absorber, low (LG) and high (HG) granularity readout. Right: the perpendicular profiles of the LG and HG     layers and the detail on the right end schematically shows the granularity of a single HG cell.}
\label{fig:strawman}
\end{figure}

The current design to meet these requirements is of a sampling calorimeter with 24 layers of 3.5 mm thick tungsten ($\sim$1 $X_0$) as absorber and Si sensitive layers of two kinds: low granularity (LG) Si PADs ($\sim$1 cm$^2$) and high granularity (HG) pixel ($\sim$1 mm$^2$). The electromagnetic section of the FoCal - depicted as the strawman design of Fig. \ref{fig:strawman} - is placed at $\sim$7 m of the interaction point, outside the magnet. 
\subsection{Performance of direct photon measurement}
\label{subsec:dirphoperf}

The current design has been simulated using the ALICE analysis framework - based on ROOT, PYTHIA and GEANT3 \cite{aliroot}- to evaluate its performance in the measurement of direct photons. The main challenge of this measurement is the extremely high background of  decay photons, the vast majority of them coming from $\pi^0$ decays. Consequently the success of this measurement depends on the efficient $\pi^0$ identification to enable the  rejection of their decay products.

The $\pi^0$ identification can be achieved either by their invariant mass reconstruction or by the shape of their reconstructed electromagnetic showers. In the first case, the two photons have enough spatial separation to allow their electromagnetic showers to be reconstructed in two separated clusters, from which an invariant mass can be calculated, and in the case of this being consistent with the $\pi^0$ mass both clusters (photons) are rejected. The design with HG layers can provide a two-shower separation distance of only a few mm. 

\begin{figure}
	\centering
	\subfloat{\includegraphics[width=.45\textwidth]{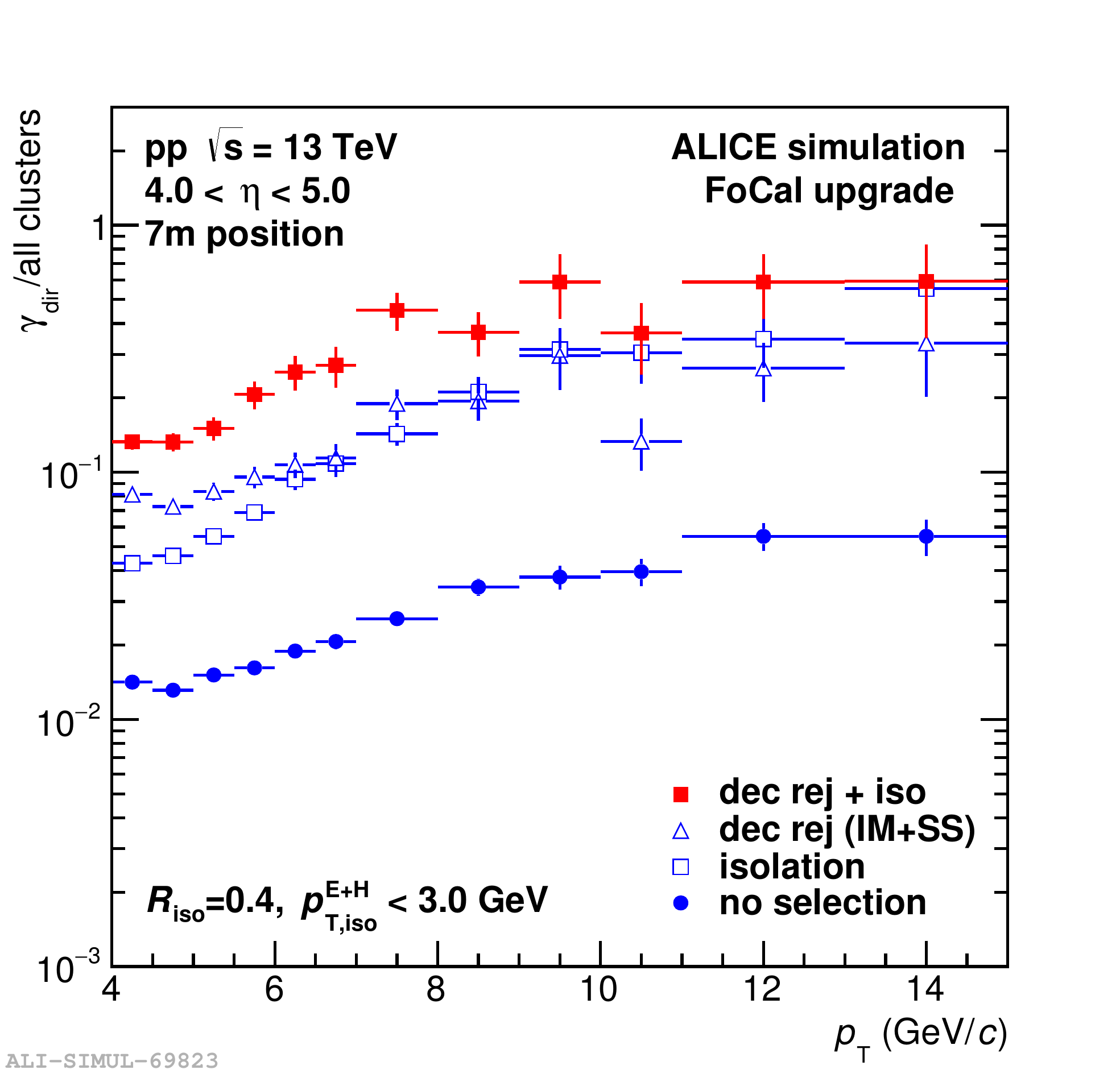}}
	\hspace{0.05\textwidth}
	\subfloat{\includegraphics[width=.45\textwidth]{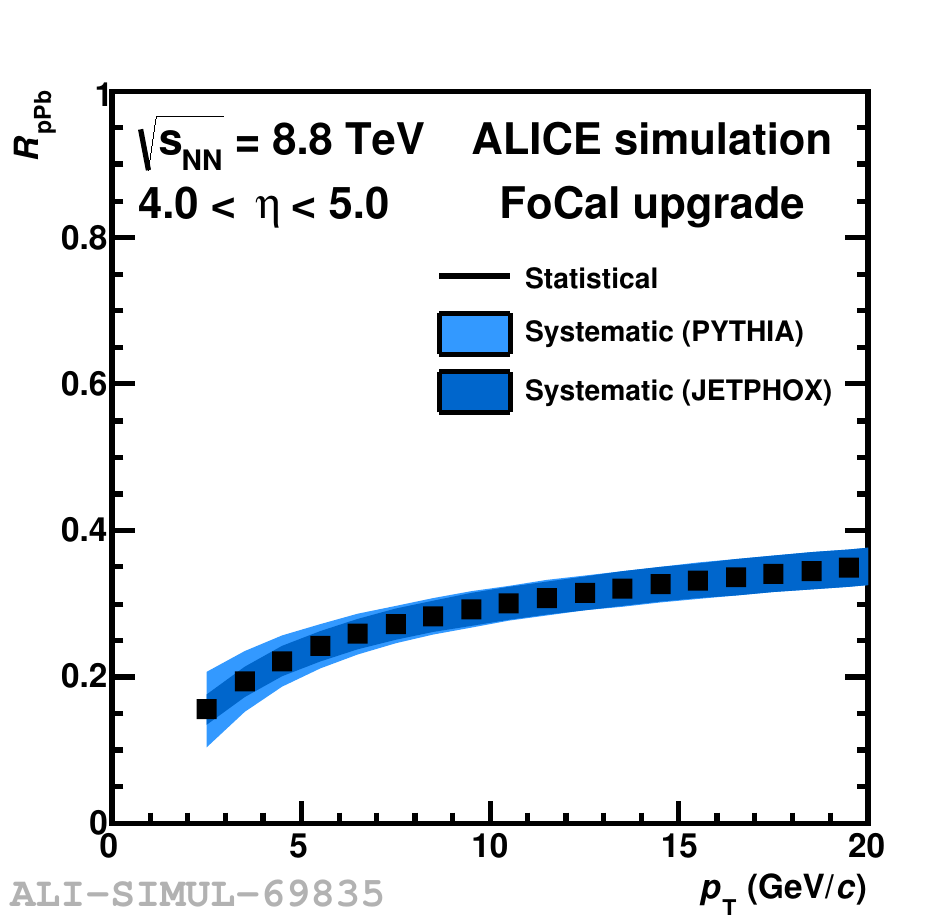}}
	\caption[]{Direct photon measurement performance. Left: rejection of background photons after the application of each analysis technique, decay rejection (Invariant Mass and Shower Shape), isolation and their combination. Right: modification factor of direct photon measurement (design with HG layers).}
\label{fig:rejection}
\end{figure}

The case of shower shape method is necessary when the $\pi^0$ energy is higher than a certain limit which boosts the decay photons to be so close spatially that their showers cannot be reconstructed separately, even with the resolution of a few mm. In this case, an invariant mass calculation is not possible, however the overlapped shower shape is very different from that of a single electromagnetic shower. This effect can be evaluated given that the calorimeter has enough granularity to reveal the details of the reconstructed cluster. The design with HG layers has a rejection power of a factor $\sim$5 higher than the option without it. The rejection power of those methods can be obtained from the left side of Fig. \ref{fig:rejection}, where the ratio of photons to all clusters is presented after each cut.
In some cases one of the decay photons cannot be reconstructed, and in such cases the other one cannot be reject using the previous methods. At these high energies, $\pi^0$s are generally associated with jets, and this kind of background can only be removed by means of isolation techniques, since direct photons should be alone (isolated), while $\pi^0$s in jets have some hadronic activity around them. For this same reason fragmentation photons can also be rejected with isolation techniques. These are the main motivations of the hadronic section of the FoCal, since the rejection power of the isolation is greatly improved with it, as shown in the left side of Fig. \ref{fig:rejection}.


After applying all rejection techniques the final results of the performance simulations is presented in the right side of Fig. \ref{fig:rejection}, as the $p_{\rm T}$ dependency of the nuclear modification factor\footnote{Calculation analogous to equation \ref{eq:rdau}.} ($R_{pPb}$). With such experimental precision the future results should be precise enough to refute or confirm scenarios such as the CGC. 

\subsection{R\&D and test beam results}
\label{subsec:testbeam}

\begin{figure}
	\centering
	\subfloat{\includegraphics[width=.40\textwidth]{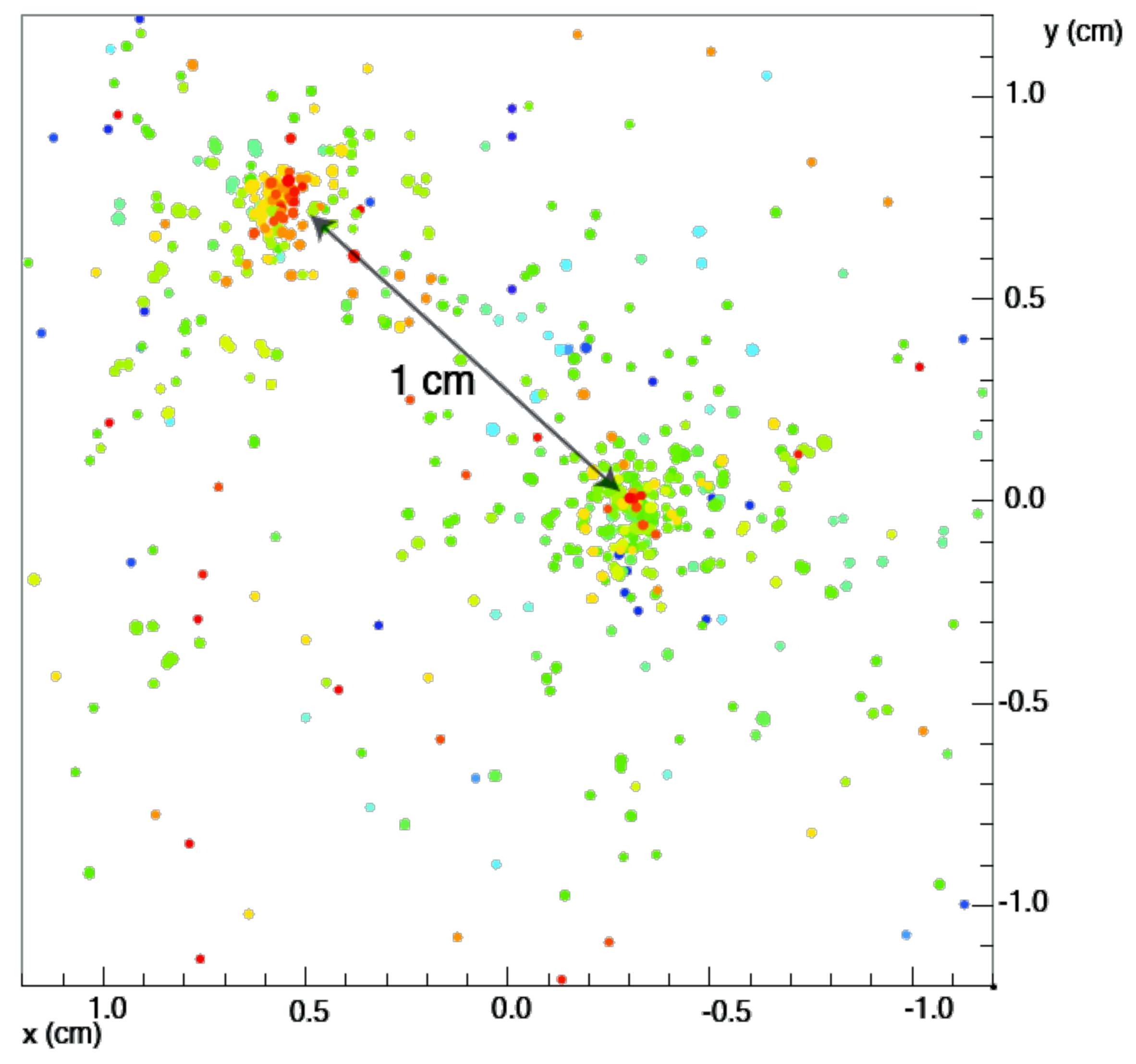}}
	\hspace{0.06\textwidth}
	\subfloat{\includegraphics[width=.42\textwidth]{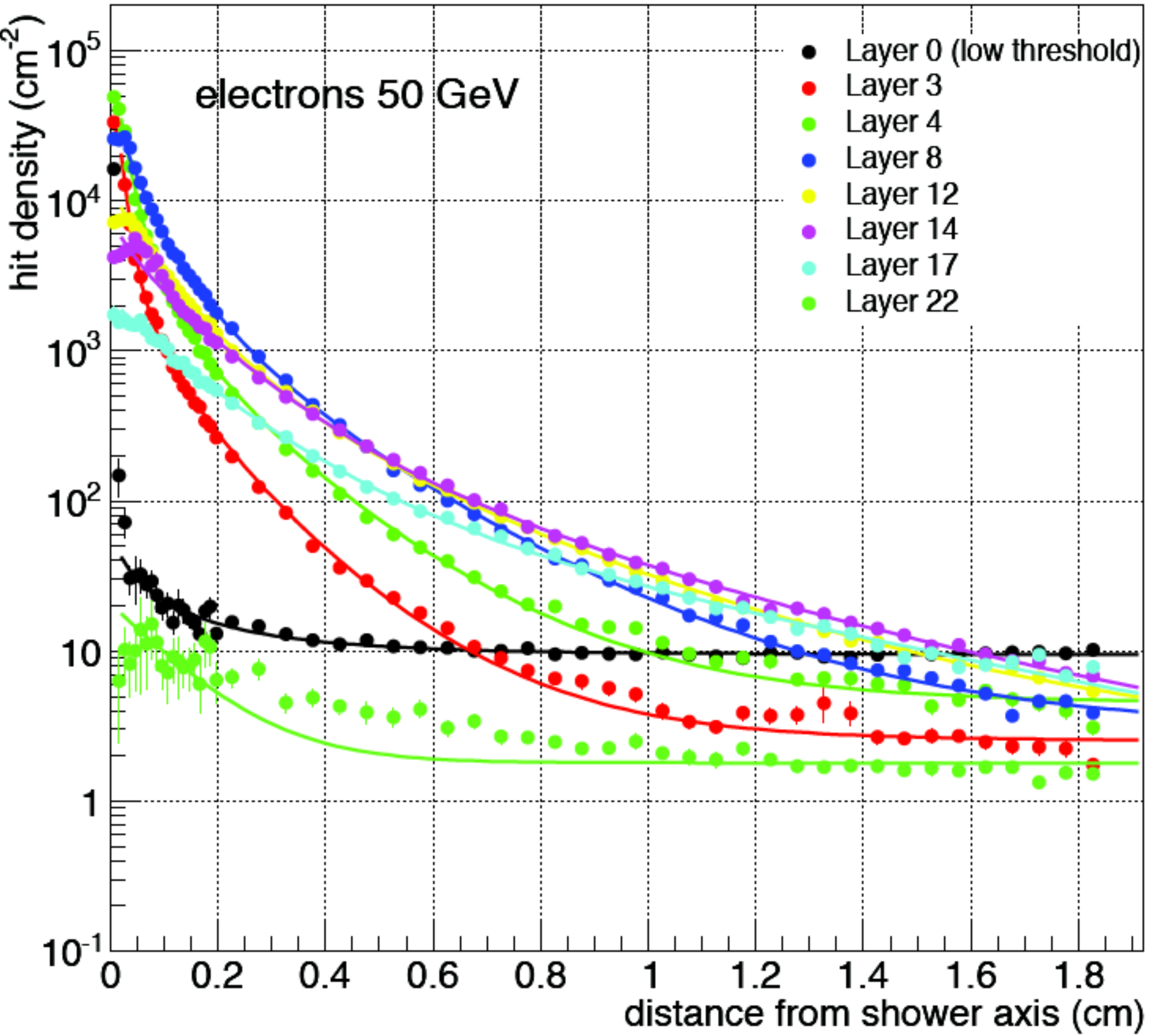}}
	\caption{Test beam results. Left: two-shower separations. Right: lateral shower profiles for a 50 GeV electron beam in selected detector layers, each corresponding to $\approx$1 $X_0$.}
\label{fig:testbeam}
\end{figure}

The FoCal collaboration is performing R\&D for LG and HG technologies, the first is likely to adopt Si-pads, while the HG layers have the Monolithic Active Pixel Sensors (MAPS) as its preferred option, due to its capability to allow very high pixel density adding little material budget. A prototype of electromagnetic calorimeter using high granularity MIMOSA23 sensors \cite{mimosa} has been built and tested. The prototype had a total number of 39
million pixels of 30 $\times$ 30 $\mu$m$^2$
in a 4 $\times$ 4 $\times$ 11 cm$^3$ volume and a Moli\`ere radius of $\sim$11 mm, although a large fraction of the shower energy is contained only in a few mm. The results, presented in Fig. \ref{fig:testbeam}, show that the prototype has been successful in reconstructing electromagnetic showers small enough to allow the separation of only a few mm distance between them.

\section{Conclusions}
In summary, the open question of the gluon saturation in the initial state hadrons and nuclei can be addressed by means of the direct photon measurement at forward rapidity, and the ALICE experiment, through the FoCal upgrade, with unprecedented spatial resolution, could provide the measurement with the necessary precision to discriminate among different theoretical models.

\acknowledgments
\begin{flushleft}
The speaker would like to thank the FoCal colleagues for the opportunity and for their support, and also to acknowledge CAPES and FAPESP for the financial support.
\end{flushleft}

\end{document}